# Chemical Equilibrium as Balance of the Thermodynamic Forces

B. Zilbergleyt,
System Dynamics Research Foundation,
Chicago, USA, E-mail: livent@ameritech.net

**ABSTRACT.**

The article sets forth comprehensive basics of thermodynamics of chemical equilibrium as balance of the thermodynamic forces. Based on the linear equations of irreversible thermodynamics, De Donder definition of the thermodynamic force, and Le Chatelier's principle, our new theory of chemical equilibrium offers an explicit account for multiple chemical interactions within the system. Basic relations between energetic characteristics of chemical transformations and reaction extents are based on the idea of chemical equilibrium as balance between internal and external thermodynamic forces, which is presented in the form of a logistic equation. This equation contains only one new parameter, reflecting the external impact on the chemical system and the system's resistance to potential changes. Solutions to the basic equation at isothermic-isobaric conditions define the domain of states of the chemical system, including four distinctive areas from true equilibrium to true chaos.
The new theory is derived exclusively from the currently recognized ideas of chemical thermodynamics and covers both thermodynamics, equilibrium and non-equilibrium in a unique concept, bringing new opportunities for understanding and practical treatment of complex chemical systems. Among new features one should mention analysis of the system domain of states and the area limits, and a more accurate calculation of the equilibrium compositions.

**INTRODUCTION.**

Contemporary chemical thermodynamics is torn apart applying different concepts to traditional isolated systems with true thermodynamic equilibrium[i] and to open systems with self-organization, loosely described as "far-from-equilibrium" area. This difference means that none of the currently recognized models allows any transition from one type of system to another within the same formalism. That's why applications of chemical thermodynamics to real objects often lead to severe misinterpretation of their status, giving approximate rather than precise results.
If a chemical system is capable of only one reaction, the reaction outcome is defined by the Guldberg-Waage's equation, based on *a priori* probabilities of the participants to interact. The situation gets complicated if several coupled chemical reactions run simultaneously. In such a system conditional rather than *a priori* probabilities constitute the Law of Mass Action (LMA). Roughly speaking, if $[R_i, R_i^\sim]$ is a dichotomial partition of the reaction space S and A is any possible reaction event on S, then defined by Bayes' theorem a conditional probability rather than an *a priori* one should be placed into LMA, as it was discussed earlier on by the author [1]. In non-ideal gases and solutions, chemical thermodynamics accounts for that implicitly, having introduced fugacities and thermodynamic activities [2]. They allow us to keep expressions for thermodynamic functions and equilibrium constants in the same appearance, disguising the open systems under the attire of isolated entities. Another case, generally thought to be a remedy to the same problem, is Gibbs' approach to phase equilibria. It represents the system as a set of open different phase entities where the equilibrium conditions include also equality of chemical potentials in addition to the traditional couple of thermodynamic parameters [3]. Actually this method is just an enhancement to the originally poorly formulated Zeroth law of thermodynamics (for amended formulation see [4]). On the opposite side of the picture are open systems with self-organization and chaotic behavior, heavily investigated and described during last three decades. In Prigogine's approach [5], which is prevailing in the field, the entropy production is the major (if not the only) factor to define the outcome of chemical processes. Following this *modus operandi* actually means implicit reduction of thermodynamic functions to entropy. The "entropic" approach is considered by some authors to be more fundamental than the "energetic" [6,7] approach. It works well in case of "weak" reactions but is not capable to cover chemical transformations with very negative changes of free energy.

---
[i] In the following discussion the term "thermodynamic equilibrium" or abbreviation "TDE" will replace "true thermodynamic equilibrium".

We do not know any serious theory trying to cover consistently both wings of chemical thermodynamics. This work is an attempt to do so on the "energetic" basis, and offers a solution, that unifies both thermodynamics aspects with a common concept in a unique theory. The preliminary results of this research were published in [8].

**DEFINITIONS.**

We have to define some new values and redefine some of the known values as well. Consider chemical reaction $\nu_A A + \nu_B B = \nu_C C$. Let $\Delta n_A$, $\Delta n_B$, $\Delta n_C$ be the amounts of moles of reaction participants, transformed as reaction proceeds from start to thermodynamic equilibrium. Obvious equalities follow from the law of stoichiometry

$$\Delta n_A/\nu_A = \Delta n_B/\nu_B = \Delta n_C/\nu_C. \quad (1)$$

Let's define the *thermodynamic equivalent of transformation* (TET) in the j-reaction as

$$\eta_j = \Delta n_{kj}/\nu_{kj}. \quad (2)$$

where $\Delta n_{kj}$ is the amount of moles of k-participant transformed in chemical reaction in j-system on its way from initial state to TDE. The numerical value of $\eta_j$ holds information of the system's $\Delta G_j^0$ and initial composition. We will use it for quantitative description of the chemical systems' composition. The above relations are strictly applicable, e.g., to reactions of species formation from elements.

De Donder [9] introduced the reaction coordinate $\xi_D$ in differential form as

$$d\xi_D = dn_{kj}/\nu_{kj} \quad (3)$$

with the dimension of mole. We re-define the reaction coordinate as

$$d\xi_Z = dn_{kj}/(\nu_{kj}\,\eta_j), \quad (4)$$

thus turning it into a dimensionless marker of equilibrium. The reaction extent $\Delta\xi_Z$ is defined as a difference between running and initial values of the reaction coordinate; obviously, the initial state is characterized by $\Delta\xi_Z=0$ while in TDE $\Delta\xi_Z=1$. This new feature allows us to define a system deviation, or shift from equilibrium in finite differences

$$\delta\xi_Z = 1 - \Delta\xi_Z. \quad (5)$$

The shift sign is positive if reaction didn't reach the state of TDE, and negative if it was shifted beyond it. In the initial state, reaction shift $\delta\xi_Z=1$ and $\delta\xi_Z=0$ in TDE. The above quantities, related to reaction coordinate, provide a great convenience in equilibrium analysis. The new reaction extent is linked to the value defined by equation (3) as

$$\Delta\xi_Z = \Delta\xi_D/\eta_j. \quad (6)$$

Further on we will use exclusively $\xi_Z$ omitting the subscript. In writing we will retain $\Delta_j$ for reaction extent and $\delta_j$ for the shift.

One of the pillars of this work is *thermodynamic force*; the author accepts Galileo's general concept of force as a reason for the changes in a system, against which this force acts [10]. Thermodynamic force (TDF) as a moving power of chemical transformations was introduced by De Donder [9] and was incorporated in chemical thermodynamics as a thermodynamic affinity

$$A_j = -(\delta\Phi_j/\delta\xi_j)_{x,y}, \quad (7)$$

where $\Phi_j$ stands for any of major characteristic functions or enthalpy, and x, y is a couple of corresponding thermodynamic parameters. This expression defines the internal affinity, or *eugenaffinity* of the j-reaction. Substitution of $\xi_D$ by $\xi_Z$ makes the affinity dimension the same as the dimension of the corresponding function in equation (7). It is very important for this work that affinity totally matches the definition of force as a negative derivative of potential by coordinate.

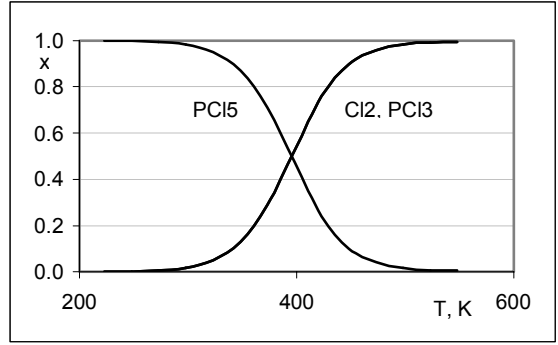

Fig.1. Equilibrium mole fractions, reaction (8), initial composition 1, 1, and 0 moles, respectively.

To illustrate major ideas and some results throughout the paper we will often use the reaction
$$PCl_3(g)+Cl_2(g)=PCl_5(g). \qquad (8)$$
This reaction is very convenient to illustrate major ideas and results of this work due to large composition changes within a narrow temperature range (Fig.1 and Table I, data obtained with HSC [11]).

Table I. Standard Gibbs' free energy changes and the thermodynamic equivalents of transformation for reaction (8) at different temperatures, p=0.1 Pa.

| T, K | 418.15 | 423.15 | 398.15 | 373.15 | 348.15 | 323.15 |
|---|---|---|---|---|---|---|
| $\Delta G^0$, kJ/mol | 5.184 | 0.901 | -3.395 | -7.704 | -12.028 | -16.365 |
| $\eta_{ji}$, mol | 0.101 | 0.240 | 0.474 | 0.713 | 0.870 | 0.950 |

**GENERAL PREMISE AND THE BASIC EQUATION OF THE THEORY.**

In our theory derivation we proceeded from the following definitions and expressions:
1. Linear equations of non-equilibrium thermodynamics with the affinities $A_{ji}$ for the internal and $A_{je}$ for the external thermodynamic force related to the j-system are represented by equation
$$v_j = a_{ji} A_{ji} + \Sigma\, a_{je} A_{je}, \qquad (9)$$
where $v_j$ is the speed of chemical reaction, and $a_{ji}$ and $a_{je}$ are the Onsager coefficients [12]. It is more constructive to put down the system's interactions in the formalism of a dichotomial section
$$v_j = a_{ji} A_{ji} + \alpha_{je} A_{je}, \qquad (10)$$
where $\alpha_{je}A_{je}$ is a contribution from the subsystem compliment [13]. Chemical equilibrium is achieved at $v_j=0$, that clearly corresponds to *equilibrium between internal and external thermodynamic forces*, causing and affecting the reaction in the j-system
$$A_{ij}^* + o_j A_{je}^* = 0. \qquad (11)$$
The dimensionless ratio $o_j = \alpha_{je}/a_{ji}$ is a reduced Onsager coefficient. One should point out that equation (11) expresses the balance between all generalized TDFs acting against the j-system; its first term is the bound affinity equal to the shifting TDF [14]. Asterisks refer values to chemical equilibrium.

2. De Donder's expression (7) for thermodynamic affinity.
3. Le Chatelier's principle. To use it, we suggest linearity between the reaction shift from TDE and external TDF ($F_{je}$) causing this shift to be

$$\delta_j = -(1/\alpha_j)F_{je}, \tag{12}$$

where $\alpha_j$ is just a proportionality coefficient, and the minus sign says that the system changes its state to decrease impact of the external TDF. Recall that $F_{je}$ is expressed in energy units; because $\delta_j$ has no dimension, the dimension of $\alpha_j$ should also be energy.

According to Le Chatelier's principle, state of the chemical system shifts from TDE until the bound affinity gets equal to the TDF to minimize or nullify its impact, i.e. $\alpha_j \delta \xi_j^* = o_j A_{je}^*$. We will place this substitution and $A_{ji}^* = (\Delta\Phi_j/\Delta_j)_{x,y}^*$ into the condition of chemical equilibrium (11), and after multiplying both sides by $\Delta_j$ we obtain

$$-\Delta\Phi_j^*(\eta_j, \delta_j^*)_{x,y} - \alpha_j \delta_j^* \Delta_j^* = 0. \tag{13}$$

This is the basic equation of the new theory. In an isolated system with $F_{je} = 0$ we have its reduced form, which is merely the traditional expression for equilibrium. Equation (13) is a typical logistic map $f(\delta_j) = \alpha_j \delta_j^*(1-\delta_j^*)$ []. It describes chemical equilibrium *in chemical systems interacting with their environment*; its reduced form is related to the TDE *of chemical reactions isolated from their environment*. It covers all virtually conceivable systems and situations, and, as we show later on, its second (parabolic) term causes a rich variety of behavior up to chaotic states.

## THE BASIC EQUATION OF STATE OF THE CHEMICAL SYSTEM AT CONSTANT PRESSURE AND TEMPERATURE.

In this case the characteristic function is Gibbs' free energy. With relation (5), equation (13) is

$$-\Delta G_j(\eta_j, \delta_j^*) - \alpha_j \delta_j^*(1-\delta_j^*) = 0, \tag{14}$$

or

$$-[\Delta G_j^0 + RT\ln\Pi_j(\eta_j, \delta_j^*)] - \alpha_j \delta_j^*(1-\delta_j^*) = 0. \tag{15}$$

Now we have a general equation for chemical equilibrium at constant p and T. It is obvious that at $\delta_j^* = 0$ this equation will reduce to the traditional $\Delta G_j^* = 0$. We will use it in a slightly different form. The dimension of $\alpha_j$ is energy, it may be interpreted as $\alpha_j = RT_{alt}$ with the second factor having dimension of temperature, an *alternative temperature*. Also, $\Delta G_j^0 = -RT\ln K$, or $\Delta G_j^0 = -RT\ln\Pi_j(\eta_j, 0)$. Being divided by RT, equation (14) changes to

$$\ln[\Pi_j(\eta_j, 0)/\Pi_j(\eta_j, \delta_j^*)] - \tau_j \delta_j^*(1-\delta_j^*) = 0, \tag{16}$$

where $\tau_j = T_{alt}/T$. We call it *reduced chaotic temperature*. This logistic equation, by analogy with the Verhulst model of population growth [16], includes shift $\delta_j^*$ as a parameter of state, $\tau_j$ as a "growth" parameter, and $\Pi_j(\eta_{kj}, 0)/\Pi_j(\eta_{kj}, \delta_j^*)$ is a reverse value of relative "chemical population" size – a ratio of the concentration function value under external impact, to the same ratio for the isolated system (the so-called maximum population size, or capacity of the isolated system). Parameter $\tau_j$ defines the "growth" of deviation from TDE; like in the Verhulst model, its numerator depends on external impact on the system (the "demand for prey" in populations [17]) while the denominator (RT) is a measure of the system resistance to changes.

## THE DOMAIN OF STATES OF THE CHEMICAL SYSTEM.

Parameter $\tau_j$ plays a critical role in the fate of dynamical systems, controlling their evolution from total extinction to bifurcations and chaos. The dependence between the reduced chaotic temperature

$\tau_j$ and the solutions to equation (16), expressed in terms of $\delta_j^*$, is known as the bifurcation diagram. In the case of a chemical system this diagram represents its domain of states. For example, bifurcation diagram for the system with reaction (8) at constant p and T is shown in Fig.2.

It is commonly accepted in the population growth theory that $0<\delta_j<1$. Unlike populations, chemical equilibrium may experience shifts to both ends, towards reactants or products; therefore it makes sense also to admit $\delta_j<0$. To illustrate this statement, the two-way bifurcation diagram with the shifts from TDE towards the initial mixture and towards the exhausted reacting mixture is shown in Fig.3. The state diagram has 4 clearly distinguishable areas, typical of bifurcation diagrams. Three out of them, having a specific meaning for chemical systems are shown in Fig.2 and Fig.3. First follows the area with zero deviation from TDE, where the curve rests on the abscissa. In this area true thermodynamic equilibrium is a strong point attractor with $\delta_j^*=0$ for all iterates: chemical equilibrium as a display of TDE totally fits itself as a display of the thermodynamic force balance. The second is the area of the open equilibrium (OPE) where the basic equation still has only one solution $\delta_j^*\neq 0$. The domain curve in both areas is the locus of single solutions to equation (16) where the iterations converge to fixed points, that is after sufficient iterations $\delta_{j(n+1)}^* = \delta_{jn}^*$ [4]. When the single solution becomes unstable, the bifurcations area with multiple values of $\delta_j^*\neq 0$ and multiple states comes out. It smoothly heads to chaos (the last, 4th area of the diagram, not shown) with increase of $\tau_j$. The magnitude of $\tau_j$ in the chemical system designates the system's position in its domain of states and defines its shift from TDE.

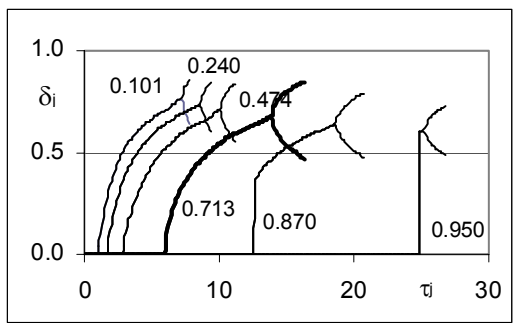

Fig.2. System domain of states, reaction (8). The numbers represent $\eta_j$ values.

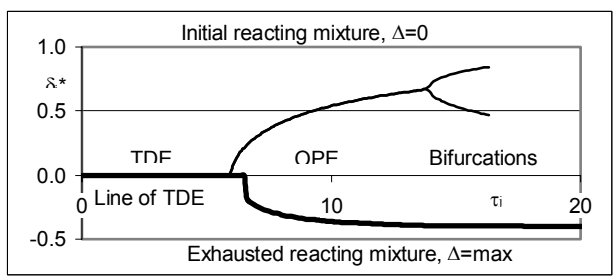

Fig. 3. Two-way diagram of states, reaction (8) at 373.15 K.

Interestingly enough, the area limits - $\tau_{TDE}$, $\tau_{OPE}$ and $\tau_{B2}$ (the limit of the period 2 bifurcations area, B2 in Fig.4) are unambiguously depending on $\Delta G_j^0$ (Fig.4). In systems with "strong" reactions ($\Delta G_j^0 \ll 0$) the most typical are the TDE and open equilibrium areas, for "weak" reactions (organic and biochemical systems) the bifurcations area may be of more importance. The limit value of $\tau_{TDE}$ is unity when $\eta_j$ tends to zero. We didn't find bifurcations in the $\delta_j^* <0$ quadrant.

The least expected and the most unusual result of the new theory is that the *TDE area is not a point but may be stretched out pretty far towards the open systems with $\tau_j >1$, up to a certain critical value of the reduced chaotic temperature*. Being unaware of any experimental proof of it, we have found some analogies using traditional way. Fig.5 shows the results of thermodynamic simulation for the equilibrium reacting mixture in the reaction of the double oxides nCaO·mRO and nBaO·mRO with sulfur, carried out at T=298 K on homological series of double oxides varying RO

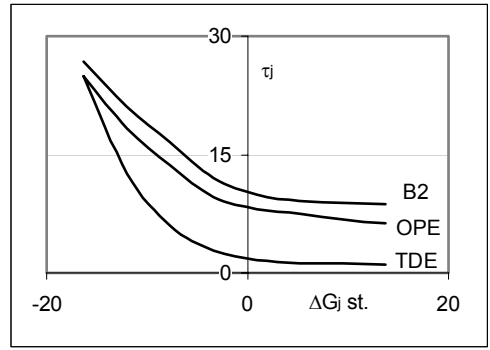

Fig.4. The area limits on the diagram of state for the system with reaction (8).

$$nMeO \cdot mRO + S <\!-\!> MeS + MeSO_4 + mRO. \qquad (17)$$

The second oxide of the nMeO·mRO couple doesn't react with sulfur at given temperature, just restricting the reactivity of MeO (RO stands for the "restricting oxide"). The abscissa on Fig.5 is reduced by RT negative Gibbs' free energy of the double oxide formation from the oxides per mole of CaO/BaO. One can see that points on abscissa in Fig.5 are protruding away from the zero point in both cases and end up with a jump like transition from the unobstructed reactivity of pure CaO/BaO and within some double oxides ($\delta_j^*=0$) to their total inertness in the double oxides located to the right of the jump point ($\delta_j^*=1$). One can get the similar information from the state domains at the same temperature as shown in Fig.5 for CaO-S ($\eta_j=0.885$, little less than the calculated value to split the curves on the graph) and BaO-S ($\eta_j=0.95$). Such a feature is typical for some double oxides at certain temperatures. The domain has the similar feature shaped by solutions to equation (16).

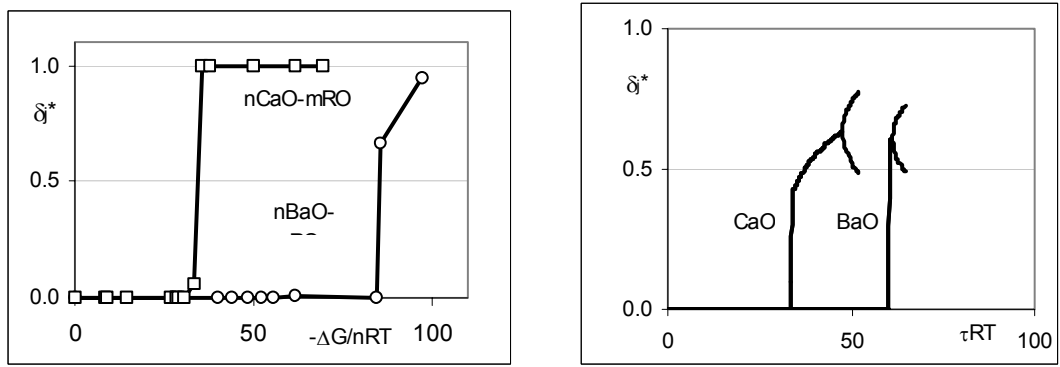

Fig.5(left). Correlation $\delta_j^*$ and $\Delta G_f^0/nRT$, reactions of nCaO·mRO/nBaO·mRO with S, 298 K.
Fig.6(right). Domains of states, (CaO+S) and (BaO+S), $\delta_j^*$ vs. $\tau_j RT$, kJ/m, 298.15 K.

The similarity between the pictures in Fig.5 and Fig.6 is quantitative: the value of $(-G_f^0/nRT)$ was taken in Fig.5 as TDF, while in Fig.6 the external force is represented by the numerator of $\tau_j$, proportional to TDF (equation (12)). Nevertheless bifurcation diagram is able to predict that kind of transitions.

**THE PROOF OF THE THEORY PREMISES.**

The only new suggestion we used to derive the basic equation is expression (12); now we will show its reasonability. As it was mentioned above, in chemical equilibrium the reaction affinity mirrors the external TDF. The graphs in Fig.7 were plotted for some simple cases based on the calculations

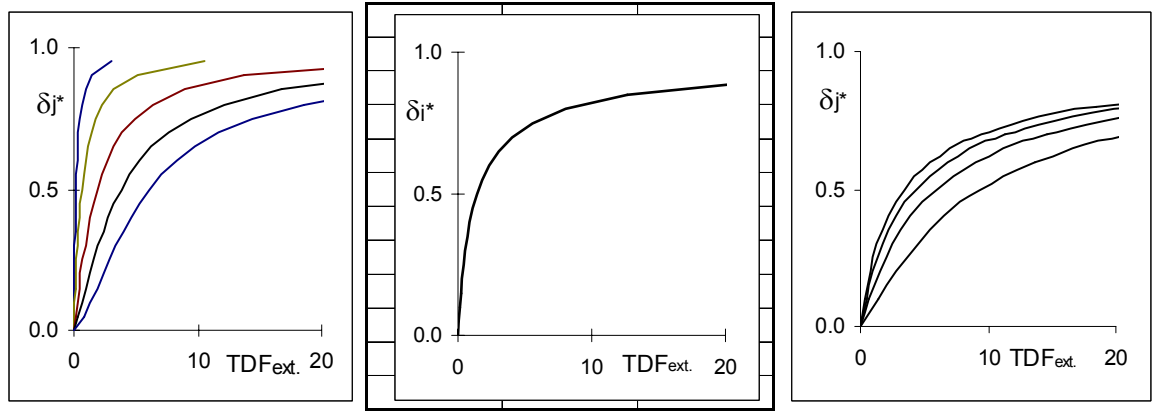

Fig.7. Shift of some simple chemical reactions from true equilibrium $\delta_j$ vs. dimensionless shifting force. Reactions, left to right: A+B=AB ($\eta$=0.1, 0.3, .., 0.9), A+2B=AB$_2$ ($\eta$=0.1, 0.2, 0.3,.., 0.9), 2A+2B=A$_2$B$_2$ ($\eta$=0.1, 0.2, 0.3, 0.4).

of TDF as $\tau_j \delta_j^*$ by varying $\eta_j$ and $\delta_j$ and using the following equation

$$F_{je}/RT = \ln[\Pi_j(\eta_j, 0)/\Pi_j(\eta_j, \delta_j^*)]/\Delta_j^*. \qquad (18)$$

In many cases the curves may be extrapolated by a straight line with the tangent values deviation ~(5-10)% up to $\delta_j^*$=(0.4–0.6). Fig.8 is related to the group of (MeO·RO+S) reactions, and the simulation was carried on as described in the previous chapter. The difference between the curve slopes for CaO·RO in Fig.5 and Fig.8 is due to the different values taken as arguments to plot the curves. Linear dependence of $\delta_j^*$ on TDF in Fig.8 is without any doubts. The restricting oxides for simulation were, in order as they follow as the dots on graphs, SiO2, Fe2O3, TiO2, WO3, and Cr2O3.

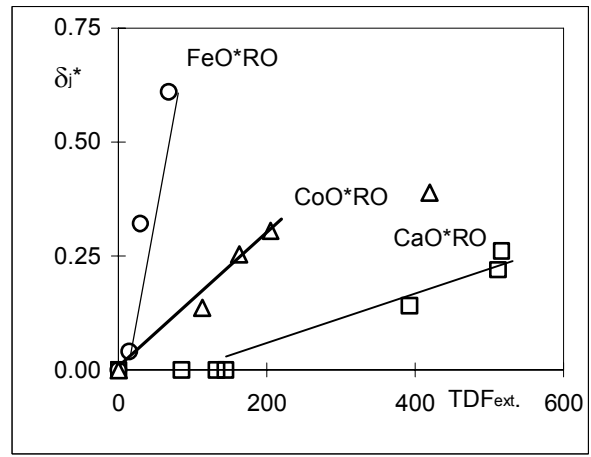

Fig.8. Dependence of $\delta_j^*$ on external force $(-\Delta G_f^0/\Delta^*)$, kJ/mol, 298.15K, reaction (17), simulation results (HSC). Points on the graphs correspond to various RO.

The above observations are proving the premise of the theory, and are closely related to the problem of finding the $\tau_j$ value for practical needs.

**AREA LIMITS AND CHARACTERISTIC REDUCED CHAOTIC TEMPERATURE.**

The new theory of chemical equilibrium presented above covers all conceivable cases – from true equilibrium to true chaos. The system location in the domain of states is controlled by the new (and only) parameter of the theory – reduced chaotic temperature. What does it change in the chemical system analysis and simulation compared to the traditional approach? If the system characteristic $\tau_j$ value falls in $[0,\tau_{TDE}]$ one should use conventional methods to calculate equilibrium composition at $\Delta G_j=0$. Else, if $\tau_j>\tau_{TDE}$ equation (16) should be used. So, we should know the area limits and characteristic $\tau_j$ value for the system in question.

The area limits may be found by direct computer simulation given initial composition and thermodynamic parameters, using iteration algorithms to solve equation (16) as described in many sources (e.g., [16]) exactly as it was done in the course of this work. On the other hand there is another time/labor saving opportunity, and the limits, $\tau_{TDE}$ and $\tau_{OPE}$ can be calculated with a good precision avoiding any simulation. For the first of them, recall that equation (16) contains 2 functions, logarithmic and parabolic. Both have at least one joint point at $\delta_j^*=0$ (Fig.9) in the beginning of the reference frame, providing for a trivial solution to equation (16) and retaining the system within the TDE area. The curves may cross somewhere else at least one time more; in this case the solution will differ from zero and number of the roots will be more than one. There is no intersection if

$$d(\tau\delta\Delta)/d\delta < d[\ln(\Pi`/\Pi^*)]/d\delta. \quad (19)$$

This condition leads to a universal formula to calculate TDE limit as

$$\tau_{TDE}=1+\eta_j\Sigma\,[\nu_{kj}/(n^0_{kj}-\nu_{kj}\eta_j)]\,, \quad (20)$$

where $n^0_{kj}$– initial amount and $\nu_{kj}$– stoichiometric coefficient of k-participant in j-system. We offer the reader to check its derivation.

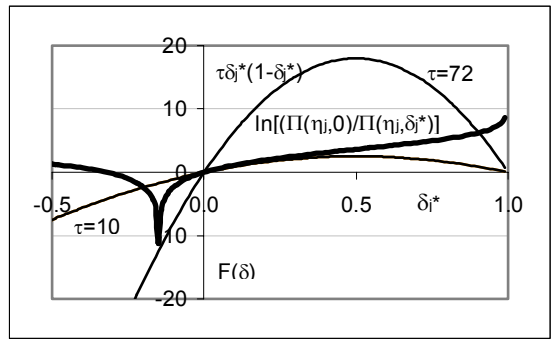

Fic.9. The terms of equation (16) calculated for reaction (8), $\eta_j=0.87$ (T=348.15K).

Though the area with $\delta_j^*<0$ is more complicated, formula (20) is still valid in cases when the system gets exhausted by one of the reactants before the minimum of the logarithmic term occurs. In case of reaction A+B=C with initial amounts of participants, corresponding to 1, 1, and 0 moles, formula (20) may be simplified as

$$\tau_{TDE} = (1+\eta_j)/(1-\eta_j). \qquad (21)$$

Fig.10 shows the comparison between values of $\tau_{TDE}$ obtained by iterative process and the calculated by formulae (20) and (21), reaction (16), in dependence on $\eta_j$. The OPE limit physically means the end of the thermodynamic branch stability where the Liapunov exponent value changes from negative to positive, and the iterations start to diverge. If the logistic equation (16) is written in the form of

$$\delta_{j(n+1)}^{*} = f(\delta_{jn}^{*}), \qquad (22)$$

the OPE limit can be found as a point along the $\tau_j$ axis where the $|f`(\delta_{jn}^{*})|$ value changes from (-1) to (+1) [4]. As of now, we do not have ready formula for this limit and would recommend finding it by iterative calculation the $\delta_j^{*}$-$\tau_j$ curve at $\tau_j > \tau_{TDE}$.

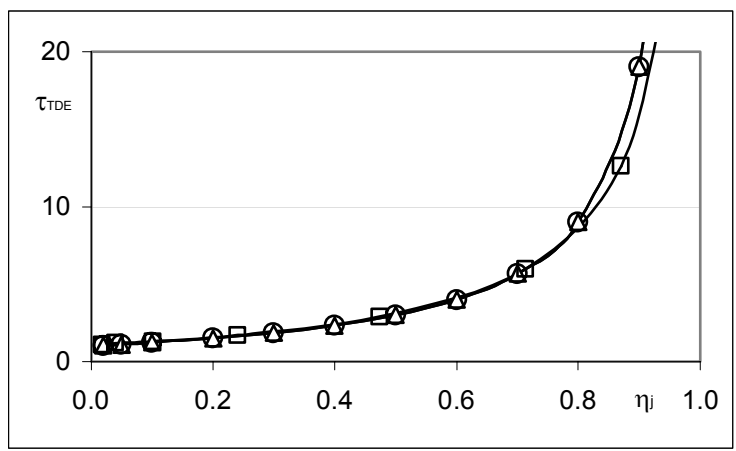

Fig.10. Calculated and simulated values $\tau_{TDE}$ vs. $\eta_j$. Series o, Δ and □ represent results calculated by equation (20), equation (21) and simulated for reaction (8) correspondingly.

The real meaning of the OPE limit is much deeper – it represents the border between the probabilistic kingdom of classical chemical thermodynamics at TDE and "close to equilibrium", on one side, and the "wild" republic of the "far-from-equilibrium" chemical systems on the other.

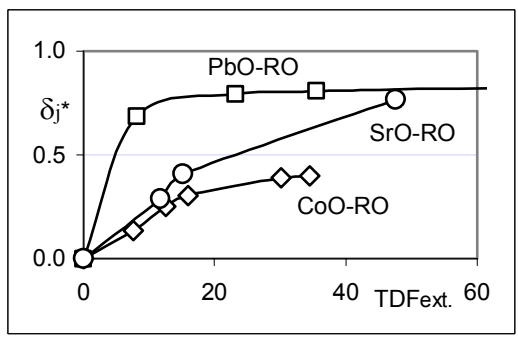

Fig.11. Shift vs. TDF in homological series of double oxides, reaction (17), HSC simulation.

There are several different ways to find $\tau_j$ within the frame of phenomenological theory. We have already touched one of them, based on the bound affinity where the sought value can be found directly from equation (18)

$$\tau_j = \ln[\Pi_j(\eta_j, 0)/\Pi_j(\eta_j, \delta_j^*)] / [\delta_j^*(1-\delta_j^*)]. \qquad (23)$$

In a certain sense it is better to find $\tau_j$ as an average of the curve tangents on the graph like in Fig.6. An alternative method consists in finding the equilibrium composition and the appropriate $\eta_j$ and $\delta_j^*$ values in the homological series by varying the external TDF. We have already described this method (see Fig.8), additional illustration to it is given in Fig.11. We have also explored a method of traditional equilibrium calculations with artificial assignment of non-unity coefficient of thermodynamic activity to any system participant. Such an approach means a restriction on the reacting ability of this participant and is based on the following reasoning. It was already mentioned that in the current paradigm interaction with the environment is accounted by means of excessive

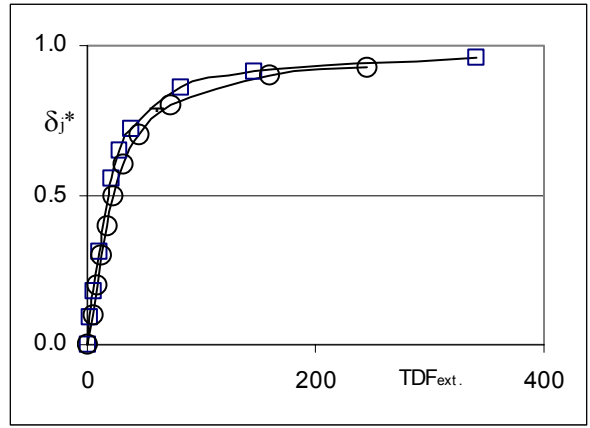

Fig.12. Force-shift graphs for reaction (27), 1000K, o and □ are related to dimensionless TDF as $\ln[\Pi_j(\eta_j, 0)/\Pi_j(\eta_j, \delta_j^*)]/\Delta_j^*$ and $(-\ln\gamma_{kj})/\Delta_j^*$ correspondingly.

functions and activity coefficients. The equilibrium condition in this case is

$$\Delta G_{rj}^* + RT \ln\Pi\gamma_{kj} = 0, \qquad (24)$$

where powers of stoichiometric coefficients are omitted for simplicity. Comparison between the reduced by RT equation (24) and equation (16) leads to the following relation between the reduced chaotic temperature and activity coefficients

$$\tau_j = (-\ln \Pi\gamma_{kj})/ [\delta_j^*(1-\delta_j^*)], \qquad (25)$$

or, in the simplest case of one coefficient per system, to

$$\delta_j^* = (1/\tau_j) [(-\ln\gamma_{kj})/\Delta_j^*], \qquad (26)$$

which is the exact replica of equation (12). At $\delta_j^* = 0$ we encounter ideality with $\gamma_{kj} = 1$ on the spot. For example, we carried out calculations for the reaction

$$2CoO+4S+2Y_2O_3=CoS2+CoS+SO2+2Y_2O_3 \qquad (27)$$

with a neutral diluent $Y_2O_3$ (non-reacting with sulfur at chosen temperature) substituting RO. The shift-force dependence for this reaction at 1000K and reactants taken in stoichiometric ratio is shown in Fig.12; the curves represent the external TDF in two different expressions. Their coincidence doesn't need any comments. So, equilibrium simulation with varying fictitious activity coefficients gives us the $\delta_j^*$ values in juxtaposition with appropriate $\gamma_{kj}$.

No surprise that parameter $\tau_j$ took a great deal of attention in this work – who knows the $\tau_j$ value, rules the chemical system. The major feature, as we see it at the moment, is that if the characteristic value of $\tau_j$ falls in $[0, \tau_{TDE}]$ one has to use conventional equilibrium conditions rather than equation (16). For instance, the values of $\tau_j$ and $\tau_{TDE}$ for the system with reaction (16) at p,T=const and initial reactant amounts (1, 1, 0) are juxtaposed in Fig.13. In this example, the characteristic $\tau_j$ value was found as average for the linear part of the force-shift curve similar to the curves plotted in Fig.7; it falls within the TDE limit. The area of linearity is matching the loosely defined "close to equilibrium" region, and the TDE approximation is good enough there for the chemical system analysis. However, we cannot offer a perfect universal method to calculate $\tau_j$ for any reaction.
Needless to say that prior to finding $\tau_j$ one has to find $\eta_j$ for the reaction in question at given temperature. It can be done by any simulation method for thermodynamic equilibrium (at $\delta_j^*=0$).

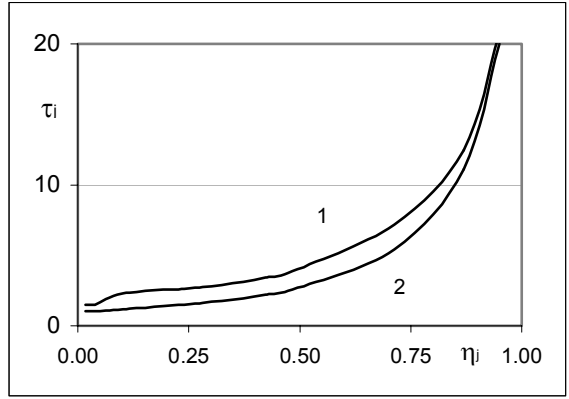

Fig.13. A correlation between $\tau_{TDE}$ (1) and characteristic $\tau_j$ (2) vs. $\eta_j$, reaction (8).

**GIBBS' FREE ENERGY OF THE CHEMICAL SYSTEM.**

The system's Gibbs' free energy change in differential form follows from equation (13) as
$$dG^*_j/RT = dG_{rj}/RT - \tau_j\,\delta_j\,d\Delta_j, \tag{28}$$
where $dG_{rj}$ is traditional differential of the reaction Gibbs' free energy. Integration of equation (28) with substitutions $G = G/RT$, $d\Delta_j = -d\delta_j$ and neglecting the integration constant gives
$$G_j^* = G_{rj}^* + \tau_j(\delta_j^*)^2/2, \tag{29}$$
or
$$G_j^* = \Sigma(n_{kj}^*)\,\mu_k^*/RT + \Sigma(n_{kj}^*)\ln\Pi_j(\eta_j,\delta_j^*) + \tau_j(\delta_j^*)^2/2. \tag{30}$$
It is common to equate $\mu^0_k$ to $\Delta G^0_{kf}$, which is related to species formation from elements, and finally we obtain an expression for system's Gibbs' free energy, reduced by RT
$$G^*_j = \Sigma(n_{kj}^*)(\Delta g^0_k) + \Sigma(n_{kj}^*)\ln\Pi_j(\eta_j,\delta_j^*) + \tau_j(\delta_j^*)^2/2. \tag{31}$$
It also belongs to the class of logistic equations but this time with positive feedback; its solutions lead to bifurcation diagram as shown in Fig.14. Area limits in this diagram are the same as found earlier for reaction (8). Obviously, the TDE area is equipotential; the system's equilibrium state and Gibbs' free energy are independent on the external impact. Though one can see well-pronounced fork bifurcations, the gaps between $G_{max}$ and $G_{min}$ are very small, averaging only 4.3 % of the larger value. For reaction (8) it seems like the fork's opposite energy levels are nearly degenerated and the system can easily switch between them. For instance, that may create a kind of a frame for the system's chemical oscillations under the influence of a non-periodic external force (for example, see [19]).

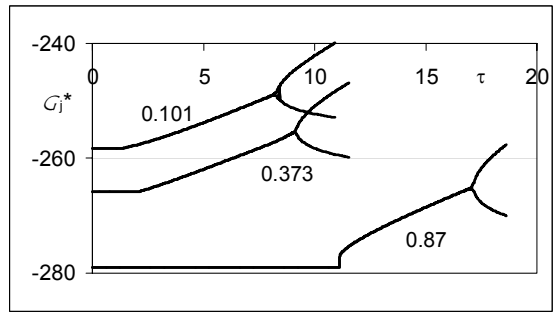

Fig.14. Reduced Gibbs' free energy vs. $\tau_j$, reaction (8). Numbers at the curves show values of $\eta_j$.

**EXAMPLE OF THE EQUILIBRIUM CALCULATIONS.**

Now we will show how the basic equation (16) works in the pre-bifurcations areas using more complicated reaction, namely

$$2CoO \cdot RO + 4S = CoS_2 + CoS + SO_2 + 2RO \qquad (32)$$

at p=0.1 Pa, T=1000K, and initial mole amounts of 1 for CoO (or CoO·RO) and of 2 for sulfur. The value of $\tau$=32.61 was obtained using the fictitious activity coefficients method (see Fig12). The joint graph for this reaction is shown in Fig.15. The ascending curves represent $\Delta_j^*$ vs. $\ln[\Pi_j(\eta_j, 0)/\Pi_j(\eta_j, \delta_j^*)]$, the distance between them along abscissa is proportional to the Gibbs' free standard energy changes of CoO·RO formation from oxides. Their intersections with the descending curve, that is $\Delta_j^*$ vs. $\tau_j\delta_j^*(1-\delta_j^*)$ give the numerical values of reaction extents. As it should be, the leftmost curve (CoO) meets the parabolic term at $\Delta$=1. Comparison of the HSC simulated reaction extents with that estimated from Fig.15 is given in Table II. One can find more examples in [20].

Table II. Equilibrium values of reaction extents in homological series, reaction (32).

|  | CoO | CoO·TiO$_2$ | CoO·WO$_3$ | CoO·Cr$_2$O$_3$ |
|---|---|---|---|---|
| $(-\Delta G^0_{f(CoO \cdot RO)}/RT)$ | 0.00 | 3.77 | 6.17 | 7.2 |
| $\Delta$ simulated, HSC | 1.00 | 0.92 | 0.89 | 0.85 |
| $\Delta$ graphical ($\tau$=32.61) | 1.00 | 0.9 | 0.82 | 0.77 |

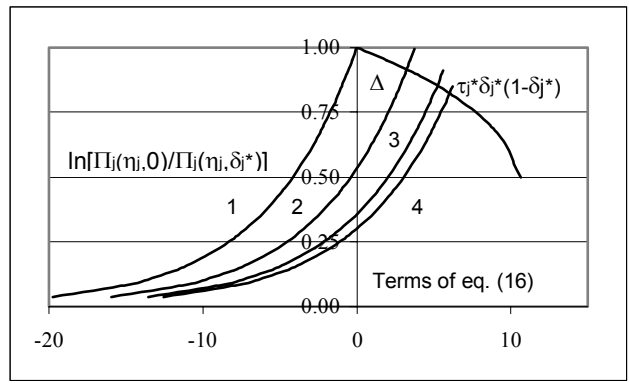

Fig.15. Reaction extent $\Delta_j^*$ vs. the terms of equation (16), reaction (32). Ascending curves - 1-CoO, 2-CoO·TiO$_2$, 3-CoO·Cr$_2$O$_3$, 4-CoO·WO$_3$.

**CONCLUSIONS.**

This work has showed explicitly that chemical equilibrium, treated as a system phenomenon, originates from the balance of internal and external thermodynamic forces which are ruling the system from within or outside. Following from such an approach the basic equation of the theory is a logistic equation, containing traditional for chemical systems logarithmic term and a new, typical for logistic equations parabolic term. Solutions to the basic equation define the domain of states of a chemical system. Chemical equilibrium matches true thermodynamic equilibrium within an initial restricted area of the domain. Within that area, the parabolic term equals to zero and the basic equation of the theory matches the traditional condition of thermodynamic equilibrium for a chemical reaction, or for an isolated chemical system. Outside the area, one has to deal with an open chemical system, where chemical equilibrium differs from classical isolated model. When the thermodynamic branch looses stability, the chemical system encounters bifurcations and chaos. The system's position in its domain of states is defined by a new parameter, the reduced chaotic temperature, which is a fraction where the numerator is proportional to the external impact on the system and the denominator reflects system's resistance against changes and merely equals to traditional RT. Application of the new theory to practice needs knowledge of that parameter; several suitable methods to find it are discussed in this work.

Major advantage of the new theory consists in extremely generalized presentation of external thermodynamic forces. Second, results of this work make it much less essential to distinguish between isolated and open systems and to draw an explicit border between them; on the calculation level the difference is automatically accounted. Introduced in this work the thermodynamics of chemical systems unites all known features of chemical systems on a common basis – from true equilibrium to true chaos.

**REFERENCES.**